\documentclass[sigconf,screen, nonacm]{acmart}
\renewcommand\footnotetextcopyrightpermission[1]{} 
\setcopyright{none} 

\AtBeginDocument{%
  }

\usepackage{tcolorbox}
\tcbuselibrary{skins,breakable,listings,theorems}

\usepackage{colortbl}
\usepackage[cachedir=minted]{minted}
\usepackage{xcolor}
\usepackage{framed}
\usepackage{booktabs}
\usepackage{pifont}
\usepackage{siunitx}
\usepackage{url}
\usepackage{tabularx}     
\usepackage{enumitem} 
\usepackage{booktabs, multirow, xcolor, graphicx}
\usepackage{ragged2e}

\usepackage[table]{xcolor}
\definecolor{OliveGreen}{rgb}{0.0,0.5,0.0} 

\newenvironment{rqbox1}{%
  \setlength{\FrameSep}{1mm}
  \setlength{\fboxrule}{0.7pt}
  \def\FrameCommand##1{\fcolorbox{black}{gray!10}{##1}}%
  \MakeFramed{\advance\hsize-\width \FrameRestore}%
  \noindent\begin{minipage}{\linewidth}
  \setlength{\parskip}{0pt}\setlength{\parindent}{0pt}
}{%

  \end{minipage}
  \endMakeFramed
}
\definecolor{rqgray}{gray}{0.90} 

\usepackage{array}
\newcolumntype{L}[1]{>{\raggedright\arraybackslash}p{#1}} 
\newcolumntype{Y}{>{\raggedright\arraybackslash}X}         
\definecolor{grayrow}{gray}{0.95}

\begin{document}

\title{The Hidden Grammar of Machines: Revealing Structural Stylometry in LLM Generated Code for Authorship Attribution}
\title{Who Wrote This JavaScript Code? Uncovering Hidden Stylometric Structures in LLM-Generated Code}
\title{Who Wrote This JavaScript Code? Uncovering Hidden Structural Patterns in LLM-Generated Programs for Authorship Attribution}

\title{The Hidden DNA of LLM-Generated JavaScript: Structural Patterns Enable High-Accuracy Authorship Attribution} 

\author{Norbert Tihanyi}
\orcid{0000-0002-9002-5935}
\affiliation{%
  \institution{Technology Innovation Institute}
  \city{Abu Dhabi}
  \country{United Arab Emirates}
}
\email{norbert.tihanyi@tii.ae}

\author{Bilel Cherif}
\orcid{0009-0006-0095-106X}
\affiliation{%
  \institution{Technology Innovation Institute}
  \city{Abu Dhabi}
  \country{United Arab Emirates}
}
\email{bilel.cherif@tii.ae}

\author{Richard A. Dubniczky}
\orcid{0009-0003-3951-1932}
\affiliation{%
  \institution{Eötvös Loránd University}
  \city{Budapest}
  \country{Hungary}
}
\email{richard@dubniczky.com}

\author{Mohamed Amine Ferrag}
\orcid{0000-0002-0632-3172}
\affiliation{%
  \institution{United Arab Emirates University}
  \city{Al Ain}
  \country{United Arab Emirates}
}
\email{mohamed.ferrag@uaeu.ac.ae}

\author{Tamás Bisztray}
\orcid{0000-0003-2626-3434}
\affiliation{%
  \institution{University of Oslo}
  \city{Oslo}
  \country{Norway}
}
\email{tamasbi@ifi.uio.no}
\authornote{Corresponding author}

\renewcommand{\shortauthors}{Tihanyi N., Cherif B., Dubniczky R. A.,Ferrag M. A., Bisztray T.}

\begin{abstract}

In this paper, we present the first large-scale study exploring whether JavaScript code generated by Large Language Models (LLMs) can reveal which model produced it, enabling reliable authorship attribution and model fingerprinting. With the rapid rise of AI-generated code, attribution is playing a critical role in detecting vulnerabilities, flagging malicious content, and ensuring accountability. While AI-vs-human detection usually treats AI as a single category we show that individual LLMs leave unique stylistic signatures, even among models belonging to the same family or parameter size. To this end, we introduce \texttt{LLM-NodeJS}, a dataset of 50,000 Node.js back-end programs from 20 large language models. Each has four transformed variants, yielding 250,000 unique JavaScript samples and two additional representations (JSIR and AST) for diverse research applications. Using this dataset, we benchmark traditional machine learning classifiers against fine-tuned Transformer encoders and introduce CodeT5-JSA, a custom architecture derived from the 770M-parameter CodeT5 model with its decoder removed and a modified classification head. It achieves 95.8\% accuracy on five-class attribution, 94.6\% on ten-class, and 88.5\% on twenty-class tasks, surpassing other tested models such as BERT, CodeBERT, and Longformer.  We demonstrate that classifiers capture deeper stylistic regularities in program dataflow and structure, rather than relying on surface-level features. As a result, attribution remains effective even after mangling, comment removal, and heavy code transformations. To support open science and reproducibility, we release the \texttt{LLM-NodeJS} dataset, Google Colab training scripts,  and all related materials on GitHub: \url{https://github.com/LLM-NodeJS-dataset}.

\end{abstract}




\maketitle

\section{Introduction}

\textit{Authorship attribution}~\cite{authorship,huang_authorship_2025}, often referred to as stylometry, is the task of determining the author of a document, text, or piece of source code by analyzing stylistic and structural features. In most existing studies on human versus AI authorship, the AI-generated category is represented by a single model~\cite{10992332,rahman_automatic_2024,nguyen_gptsniffer_2024,info15120819} or, at best, a small set of models~\cite{10.1145/3605770.3625215}, thereby limiting the robustness of the resulting methods~\cite{pan2024assessing}.

In human code attribution, typically only limited samples per author are available~\cite{caliskan-islam_-anonymizing_2015,abuhamad_large-scale_2018}, while \textit{large language models (LLMs)} allow the generation of virtually unlimited, task-specific datasets for mapping their coding style. However, because many LLMs are trained on overlapping code corpora~\cite{TheStack2022}, they may inherit a broad mixture of stylistic and structural characteristics. With the growing presence of AI-generated content online, researchers have also raised concerns about recursive training and eventual style convergence among models~\cite{Shumailov2023CurseOfRecursion,Mitchell2023DataContamination}. If such convergence occurs, model-level attribution, or even model family detection may become increasingly difficult.

Big-tech leaders predict that AI will soon generate most of our source code~\cite{Amodei2025,Nadella2025,Altman2025}. Thus, robust attribution in LLMs is urgently needed for critical domains like software forensics, malware analysis, cybercrime investigations, and academic integrity~\cite{OmanCook1989}. Recent work by Bisztray et al.~\cite{bisztray2025know} shows that different LLMs leave distinguishable fingerprints when writing C code, achieving 94.5\% accuracy in telling apart C code generated by five leading LLMs.

These findings highlight that attribution research must extend beyond the binary AI-vs-human setting to richer tasks such as model-level attribution, family-level attribution, co-authorship attribution, obfuscation-robust attribution, and malware/threat actor attribution. These tasks differ not only in dataset accessibility but also in their sensitivity to model evolution, as LLM fingerprints can change rapidly with the introduction of new models.

In \cite{bisztray2025know}, code stylometry focused on distinguishing C code among five LLM families (Claude-3.5, DeepSeek-v3.1, Gemini-2.5, GPT-4.1, LLaMA-3.3-70B). Extending this line of work, we focus on JavaScript---a flexible, dynamically typed language whose diverse styles and widespread tooling make attribution more challenging. We further investigate whether stylometric attribution is possible within the same model family (GPT-4o, GPT-4o-mini, GPT-4.1, GPT-5, gpt-oss) and explore scaling to larger classification tasks (10–20 classes). To this end, we introduce the first diverse AI-generated JavaScript dataset for authorship attribution and address the following research questions:

\begin{figure*}[ht] 
\centering
\caption{The seven-step methodology for LLM authorship attribution in JavaScript code.}
\includegraphics[width=0.98\textwidth]{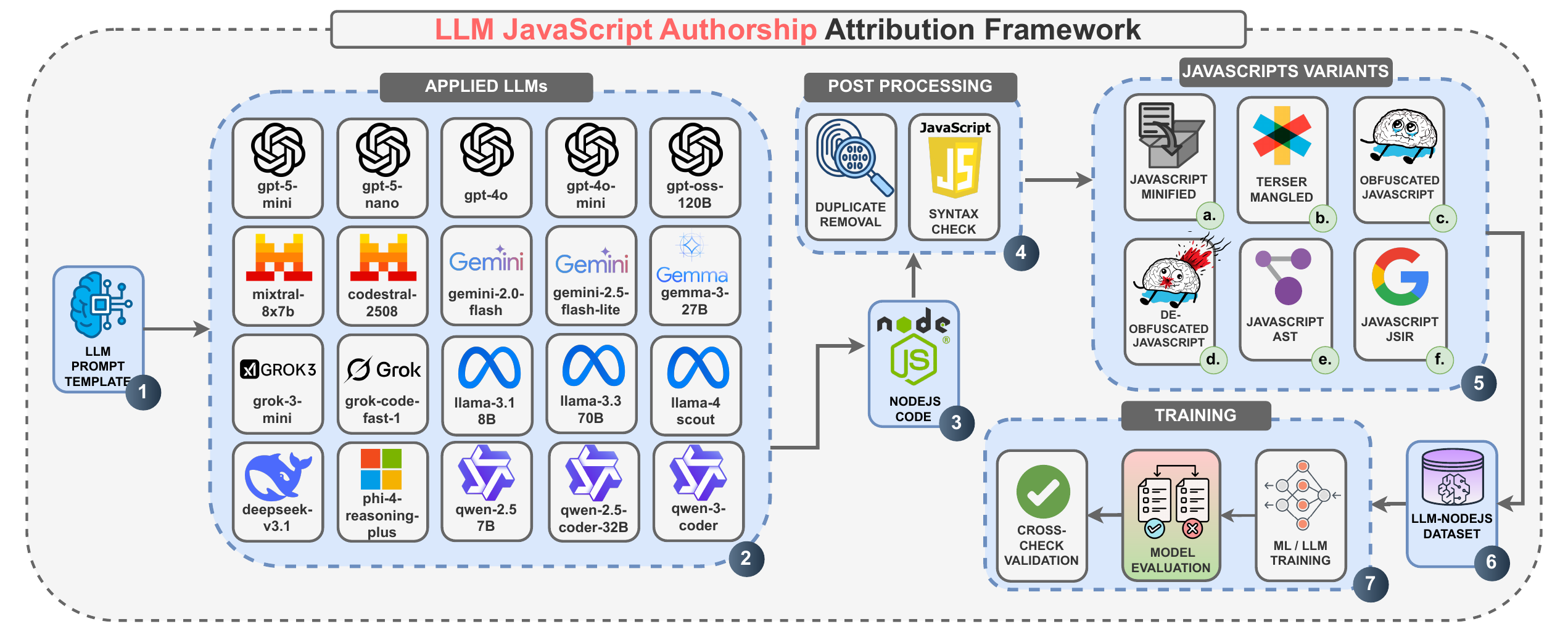} 
\label{fig:Framework}
\Description{The four-step methodology for LLM authorship attribution in C code.}
\end{figure*}

\begin{tcolorbox}[
    colback=gray!10,    
    colframe=black,     
    arc=6pt,            
    boxrule=0.7pt,      
    left=2mm, right=2mm, top=1mm, bottom=1mm
    ]

\small    

\textbf{RQ1:} Which machine learning and transformer-based approaches achieve the most robust performance for black-box, model-level JavaScript source code attribution? \\[0.5em]
\textbf{RQ2:} What signals do deep-learning methods exploit to distinguish between different models?
\end{tcolorbox}

To answer these questions, we follow the methodology shown in Figure~\ref{fig:Framework}.
On a high level, our pipeline begins with prompt creation for back-end Node.js coding tasks and code generation from 20 state-of-the-art LLMs. The outputs are syntax-checked, deduplicated, and transformed into multiple JavaScript variants. These are then used to train both traditional \textit{Machine Learning} (ML) models and transformer-based models for authorship attribution, which are subsequently evaluated and validated on an independent dataset. The detailed methodology is introduced in Section~\ref{sec:methodology}.

In this paper, we present the following key contributions:
\begin{itemize}

\item \textbf{LLM-NodeJS dataset:} We release a public dataset of $50{,}000$ syntactically correct and executable Node.js back-end programs, expanded with four additional code variants (minified, mangled, obfuscated, deobfuscated), resulting in $250{,}000$ JavaScript samples. Covering 20 state-of-the-art models and diverse web tasks, the dataset also includes two additional representations \textit{JavaScript Intermediate Representation} (JSIR) and \textit{Abstract Syntax Tree (AST)}, providing the first large-scale foundation for a wide range of JavaScript research, not limited to code authorship.

\item \textbf{Classifier evaluation:} Using the \textsc{LLM-NodeJS} dataset, we benchmark traditional ML classifiers (Random Forest, XGBoost, KNN, Logistic Regression, Linear SVM) and fine-tuned transformers (i.e., BERT, CodeBERT, Longformer), including a custom \textit{CodeT5-Authorship} model with a modified classification head and removed decoder layers. We show that JavaScript attribution is highly effective, achieving over 95\% accuracy across five models within the same GPT family (GPT-4.1, GPT-4o, GPT-4o-mini, GPT-5-nano, GPT-5-mini, GPT-oss-120b) and nearly 90\% for 20-class attribution.

\item \textbf{Robustness to code variants:} Attribution remains accurate on minified (91\%), terser-mangled (93\%), and deobfuscated (85\%) code, while obfuscation lowers accuracy due to token inflation. These results show that classifiers rely on deeper structural signals—such as AST structure and data-flow graphs—rather than shallow stylistic features explored in prior work

\item \textbf{Open science:} To support reproducibility, we release the \textsc{LLM-NodeJS} dataset along with all Google Colab notebooks and training scripts on the project GitHub page: \url{https://github.com/LLM-NodeJS-dataset}.

\end{itemize}

\begin{tcolorbox}[
    enhanced,
    title=Key Takeaway,
    colframe=black!50,
    colback=gray!10!white,
    arc=1mm,
    fonttitle=\bfseries,
    coltitle=black!50!black,
    attach boxed title to top text left={yshift=-0.50mm},
    boxed title style={
        skin=enhancedfirst jigsaw,
        size=small,
        arc=1mm,
        bottom=-1mm,
        interior style={
            fill=none,
            top color=gray!30!white,
            bottom color=gray!20!white
        }
    }
]
The key insight of our work is that individual LLMs leave distinctive, structurally grounded fingerprints, enabling reliable model-level attribution beyond traditional AI-vs-human detection. \textbf{As AI-generated code becomes standard, the key question shifts from whether it was written by AI to which AI produced it}, highlighting the need to treat the “AI-generated” category as diverse rather than uniformly representing it with one model.
\end{tcolorbox}

The rest of the paper is organized as follows. Section~\ref{sec:related_work} reviews related work. Section~\ref{sec:methodology} outlines the methods used to construct the dataset and to train the classifiers. Section~\ref{sec:experimental_results} presents experimental results, Section~\ref{sec:limitations} discusses limitations and future work, while Section~\ref{sec:conclusion} concludes the paper.

\section{Related Work}
\label{sec:related_work}

This section reviews authorship attribution methods for source code, organized by technique family. We then discuss benchmark datasets and tasks.

\subsection{Authorship attribution architectures}

\subsubsection{Traditional Machine Learning}
Authorship attribution has long been framed as supervised classification, mapping code features into vectors for classifiers such as logistic regression, naïve Bayes, SVMs, or tree ensembles~\cite{he_authorship_2024}. These methods are cheap and interpretable but rely on feature engineering and lag behind deep learning in accuracy~\cite{bisztray2025know}.

\subsubsection{Deep Learning}
Neural models learn directly from code, capturing stylistic hierarchies that manual features miss~\cite{zafar_language_2020}. Abuhamad et al.\ achieved 92.3\% accuracy on 8,903 authors in Google Code Jam with an RNN+Random Forest hybrid, also showing robustness to obfuscation~\cite{abuhamad_large-scale_2021}. Such models scale better to thousands of authors but require large labeled data.

\subsubsection{Pre-trained and Generative Models}
Pre-trained models (i.e., BERT, CodeBERT, RoBERTa) learn contextual embeddings from large code corpora, yielding state-of-the-art attribution with minimal manual features~\cite{feng_codebert_2020,guo_graphcodebert_2021,wang_codet5_2021}. Fine-tuned CodeT5 variants achieve strong accuracy even across closely related authors~\cite{howard-ruder-2018-universal}.  

Generative LLMs (e.g., GPT-4) support zero/few-shot attribution, reaching 65–69\% on hundreds of authors~\cite{choi_i_2025}. Beyond human code, Bukhari et al.\ distinguished AI- from human-written C code at 92\%~\cite{10.1145/3605770.3625215}, while Idialu et al.\ reported F1 $\approx$ 0.91 for GPT-4 vs.\ human Python~\cite{idialu2024whodunit}, though most treat AI as a single-source class. Bisztray et al.\ extended this to the AI-vs-AI setting, with a modified \emph{CodeT5} surpassing 95\% accuracy across five LLMs from different model families~\cite{bisztray2025know}.

Together, these results show both human and machine code leave detectable stylistic signatures.

\subsection{Code Representation}

\subsubsection{Feature Engineering Approaches}
Early work fingerprinted authors via lexical cues (keywords, punctuation), formatting habits, and software metrics~\cite{tihanyi_how_2024}. Caliskan-Islam et al.\ introduced the \textit{Code Stylometry Feature Set }(CSFS)~\cite{caliskan-islam_-anonymizing_2015}. While effective on human subjects, hand-crafted features degrade under obfuscation, adversarial edits, or IDE suggestions~\cite{quiring_misleading_2019,simko2018,mcknight2018}.

\subsubsection{Graph and Structural Representations}
Graphs capture structural and semantic style. ASTs encode syntax~\cite{10.1007/978-981-16-6372-7_70}, while \textit{Program Dependence Graphs (PDGs)} and \textit{Control Flow Graphs (CFGs)} highlight dependencies and control flow. Bogomolov et al.\ used AST path–based classifiers (PbRF, PbNN) to reach 98\% on curated sets but saw sharp drops on realistic corpora~\cite{bogomolov2021authorship}. Graph-based features are more robust to formatting changes than stylometry~\cite{guo_method_2022,he_authorship_2024}.

\subsubsection{Obfuscated Code Attribution}
Static stylistic features are fragile under obfuscation~\cite{quiring_misleading_2019,simko2018,mcknight2018}, but syntax and AST-based methods remain effective~\cite{caliskan-islam_-anonymizing_2015}. One Study achieved $\sim$93\% accuracy even on obfuscated code using deep learning ~\cite{abuhamad_large-scale_2018}.

\subsubsection{Execution and Binary-Level Approaches}
Dynamic analysis leverages runtime traces (system calls, memory, performance) to reveal algorithmic signatures resistant to superficial edits~\cite{wang_integration_2018}. However, sandboxing is costly and mostly limited to malware forensics~\cite{9825799,10.1145/3653973,7784595}. Attribution also extends to binaries: stylistic markers often survive compilation~\cite{rosenblum_who_2011,he_authorship_2024}, and despite compiler variability, accuracy remains above chance~\cite{10.1145/3653973}.

\subsection{Datasets, Benchmarks and Tasks}
Authorship attribution depends on diverse labeled corpora. Widely used datasets for human attribution include Google Code Jam, Codeforces, GitHub OSS, POJ-104, and BigCloneBench~\cite{caliskan-islam_-anonymizing_2015,alsulami_source_2017,abuhamad_large-scale_2018,abuhamad_large-scale_2021,bogomolov2021authorship}. Yet, classifiers designed for natural language reportedly perform poorly on source code~\cite{pan2024assessing}.  For AI vs.\ human attribution, most studies treat AI as a single-model category~\cite{10.1145/3605770.3625215,idialu2024whodunit}. Both studies focus exclusively on machine learning methods and report strong results on binary classification of AI vs. human-written C code (F1 $\approx$ 0.91 and Accuracy $\approx$ 92\%). However, their evaluation is limited to GPT-4 as the sole source of AI-generated code. The only large-scale AI vs.\ AI study is by Bisztray et al.~\cite{bisztray2025know}, who introduce the LLM-AuthorBench dataset (32k C programs). They achieve 95\% accuracy in distinguishing five LLMs, with transformers consistently outperforming traditional ML methods.

\section{Methodology}
\label{sec:methodology}
The first part of our methodology involves constructing a new supporting dataset, \texttt{LLM-NodeJS}, to address the lack of syntactically correct and labeled LLM-generated JavaScript code in existing literature.

\subsection{Dataset Creation- The LLM-NodeJS dataset}

We designed a set of $250$ complex back-end Node.js tasks inspired by real-world scenarios encountered by professional developers. Formally, let
\[
\mathcal{T} = \{ T_1, T_2, \dots, T_{250} \}
\]
denote the set of tasks. Each task $T_i$ is decomposed into a collection of smaller subtasks, $T_i = \{ t_{i1}, t_{i2}, \dots, t_{i m_i} \}$, 
where $m_i$ is the number of subtasks in task $T_i$. These subtasks represent functionally meaningful components (e.g., writing a GitHub uploader, creating a URL shortener service with a local SQLite database) and introduce substantial structural and stylistic diversity, ensuring that no two generated programs in the \texttt{LLM-NodeJS} dataset are identical.

All generated JavaScript implicitly uses strict mode and ES6 imports instead of \texttt{require}. For example, one task involved creating a JWT validator service that verifies RSA-signed tokens and returns a JSON status, comprising dozens of smaller subtasks. For each task $T_i \in \mathcal{T}$, we generated $k = 10$ independent implementations per model, yielding
\[
|\mathcal{D}_m| = k \times |\mathcal{T}| = 10 \times 250 = 2{,}500
\]
samples per model $m$. Across twenty LLMs spanning eight model families (c.f. Figure~\ref{fig:Framework}), this resulted in a total of $50{,}000$ unique programs. Only JavaScript snippets that were syntactically correct and passed \texttt{"node --check"} command were included in the dataset. To assess the effect of obfuscation on attribution accuracy, we created six variants of each sample:

\begin{enumerate}
\small
    \item \textbf{JavaScript Minified} – whitespace, comments, and formatting removed.  
    \item \textbf{Terser Mangled} - identifiers shortened systematically via Terser.  
    \item \textbf{Obfuscated JavaScript} - transformed using techniques such as control-flow flattening or string encoding.  
    \item \textbf{De-Obfuscated JavaScript} - partially restored versions improving readability through identifier recovery and reformatting.  
    \item \textbf{JavaScript JSIR} - intermediate representation for program analysis and attribution.  
    \item \textbf{JavaScript AST} - structured tree representation capturing syntax and hierarchy.  

\end{enumerate}

The dataset includes the validated original programs plus four syntactically correct variants (minified, mangled, obfuscated, deobfuscated), comprising a total of $250{,}000$ JavaScript samples, as well as two additional representations (JSIR and AST) that support a wide range of JavaScript analysis research.

\subsection{Selected Models for Dataset Evaluation}
The second step of our analysis focused on evaluating three different approaches: (i) traditional ML algorithms, (ii) transformer-based encoder-only models, and (iii) a custom model with a modified architecture. These methods were applied to the newly created dataset to assess their performance.

\subsubsection{Traditional ML Algorithms}
\label{sec:ml-js-stylometry}
As a baseline, we evaluated five different traditional ML algorithms on stylometric representations of JavaScript code, including \textit{$k$-NN, Logistic Regression, Random Forest, Linear SVM}, and \textit{XGBoost}. The pipeline includes pre-processing, feature extraction, and training. Each program was vectorized using TF–IDF over tokens with a vocabulary cap of 400, emphasizing lexical and control-flow patterns known to capture stylistic cues~\cite{caliskan-islam_-anonymizing_2015,caliskan_when_2018}. Performance was evaluated using accuracy, precision, F1 score, and training time. Since our dataset is balanced, recall does not provide additional information.

\subsubsection{Transformer-based Models }
\label{sec:transformers}

We evaluate three encoder-only architectures~\cite{vaswani_attention_2017}, including BERT, CodeBERT, and Longformer. Encoder models are typically favored for classification, while decoder models are preferred for generation with long context. 

All transformer fine-tuning experiments were optimized for NVIDIA~A100 (80\,GB) GPUs. Batch sizes, gradient accumulation, and mixed-precision settings were adjusted to fully utilize available memory while avoiding overflow. In particular, training was performed in \texttt{bf16} precision with fused AdamW optimization and cosine restarts for learning rate scheduling. For BERT, we used an effective batch size of 32 (achieved through 16 × 2 gradient accumulation), whereas Longformer leveraged its larger context window with a batch size of 32 and no accumulation.

\subsubsection{Custom CodeT5-JSA Architecture}

In ~\cite{bisztray2025know}, the authors transformed CodeT5+ (770M parameters) into an encoder-only model by removing the decoder layer and attaching a lightweight classification head (H1) consists of:
\vspace{-0.02cm}
\[
\setlength{\fboxsep}{1.5pt} 
\begin{aligned}
&\fbox{\footnotesize\parbox[c][9pt][c]{1.2cm}{\centering Linear layer}} \hspace{1pt}\rightarrow\hspace{1pt}
 \fbox{\footnotesize\parbox[c][9pt][c]{0.6cm}{\centering GELU}} \hspace{1pt}\rightarrow\hspace{1pt}
 \fbox{\footnotesize\parbox[c][9pt][c]{1.3cm}{\centering Dropout(0.2)}} \hspace{1pt}\rightarrow\hspace{1pt}
 \fbox{\footnotesize\parbox[c][9pt][c]{1.2cm}{\centering Linear layer}}
\end{aligned}
\]
For the ten and twenty class scenario the above setting has started dropping accuracy, therefore here we revised the classification head:
\vspace{-0.02cm}
\[
\setlength{\fboxsep}{1.5pt} 
\begin{aligned}
&\fbox{\footnotesize\parbox[c][9pt][c]{1.3cm}{\centering Dropout(0.2)}} \hspace{0.5pt}\rightarrow\hspace{0.5pt}
 \fbox{\footnotesize\parbox[c][9pt][c]{1.2cm}{\centering Linear layer}} \hspace{0.5pt}\rightarrow\hspace{0.5pt}
 \fbox{\footnotesize\parbox[c][9pt][c]{0.7cm}{\centering Tanh}} \hspace{0.5pt}\rightarrow\hspace{0.5pt}
 \fbox{\footnotesize\parbox[c][9pt][c]{1.3cm}{\centering Dropout(0.2)}} \hspace{0.5pt}\rightarrow\hspace{0.5pt}
 \fbox{\footnotesize\parbox[c][9pt][c]{1.2cm}{\centering Linear layer}}
\end{aligned}
\]
In both settings the last layer is followed by \texttt{Softmax} activation to produce class probabilities. Replacing the GELU activation layer with \texttt{Tanh} and introducing an initial dropout layer improved stability and overall accuracy. For example on the deobfuscated twenty class scenario accuracy jumped from 72\% to 79.94\%.

In terms of optimisation, if full FP32 is used, the training time is 958 minutes, whereas switching to bfloat16 reduces the 20-class training time to 111 minutes, with only a 1–3\% accuracy drop, which is still acceptable given the substantial training time gain.




\section{Experimental Results}
\label{sec:experimental_results}
To answer our research questions, we conducted our experiments in two stages. The first is geared towards research question 1, where we investigate which machine learning and transformer-based architectures perform best in attributing JavaScript code (along with the obfuscated versions) to the original generator model. 

Next, we aim to investigate if attribution performance holds even when there are more classes, and whether there are certain models, that are too similar to each-other to be accurately distinguished.

\subsection{Dataset diversity analysis}
Before starting the classification process, we assessed the dataset's diversity. As discussed in the methodology section, each task is posed ten times to the same model, enabling us to measure variability in the generated JavaScript programs. To quantify this diversity—both within and across model families—we used three \textit{CodeBLEU} components: \textbf{N-gram match} (lexical overlap on tokens, identifiers, and keywords), \textbf{Syntax match} (structural overlap via ASTs), and \textbf{Dataflow match} (semantic overlap via variable dependencies and data/control flow, e.g., Jaccard over dependency graphs).
Table~\ref{tab:match} reports \textit{median} similarities for GPT-4o, GPT-5, and Gemini. We summarize intra--vs.--inter model separation by the average gap $\Delta$ ($\sum$ intra-model medians - $\sum$ inter-model medians). \textit{Lower similarity indicate greater diversity in the code, whereas for $\Delta$ higher values reflect stronger discriminative power between models.}

\begin{table}[htb]
\centering
\footnotesize
\caption{Median CodeBLEU Similarities and Intra–Inter Gaps for JavaScript Programs}
\label{tab:match}
\renewcommand{\arraystretch}{1.1}
\setlength{\tabcolsep}{6pt}
\begin{tabular}{lccc}
\toprule
\textbf{Model Pair} & \textbf{N-gram Match} & \textbf{Syntax Match} & \textbf{Dataflow Match} \\
\midrule
\rowcolor{gray!17}
\multicolumn{4}{c}{\textbf{Intra-Model Comparisons}} \\
Gemini × Gemini & 0.29 & 0.58 & 0.40 \\
GPT-4o × GPT-4o & 0.29 & 0.60 & 0.46 \\
GPT-5 × GPT-5   & 0.16 & 0.60 & 0.35 \\
\midrule
\rowcolor{gray!17}
\multicolumn{4}{c}{\textbf{Inter-Model Comparisons}} \\
Gemini × GPT-4o & 0.09 & 0.49 & 0.31 \\
Gemini × GPT-5  & 0.03 & 0.43 & 0.16 \\
GPT-4o × GPT-5  & 0.01 & 0.38 & 0.11 \\
\midrule
\midrule
\textbf{Intra-Model Avg.} & 0.25 & 0.59 & 0.40 \\
\textbf{Inter-Model Avg.} & 0.04 & 0.43 & 0.19 \\
\textbf{Gap ($\Delta$)}          & \textbf{0.20} & \textbf{0.16} & \textbf{0.21} \\
\bottomrule
\end{tabular}
\end{table}

\subsubsection{Key Results}
\emph{Dataflow similarity} provides the strongest discrimination between same and cross-model pairs, representing the clearest model-specific signal at the semantic level. 
Programs compared within the same model are semantically aligned (e.g., GPT-4o~$\times$~GPT-4o median $0.46$), whereas cross-family pairs can be markedly divergent (GPT-4o~$\times$~GPT-5 median $0.11$). 
\emph{N-gram} similarity shows a comparable gap but is sensitive to surface-level edits. \emph{Syntax} yields the smallest gap, not because it lacks discriminative power, but because structural patterns are relatively stable across tasks. 

\subsubsection{Implications for attribution}
These results indicate that model attribution will be most reliable when grounded in deeper semantic and structural signals rather than surface-level cues. 
Even when stylistic features are obfuscated or lexical patterns altered, the underlying logic and structural organization of code often remain detectable. 
\emph{Dataflow similarity} thus provides the most obfuscation-resilient indicator of model identity, capturing consistent reasoning and dependency patterns. 
\emph{Syntax similarity} offers complementary stability, reflecting structural regularities that persist despite superficial transformations, while \emph{n-gram similarity}—although highly discriminative in clean conditions—degrades rapidly under lexical variation.

\subsubsection{Token length distribution}
\begin{figure}[htb]
    \centering
    \includegraphics[width=1\linewidth]{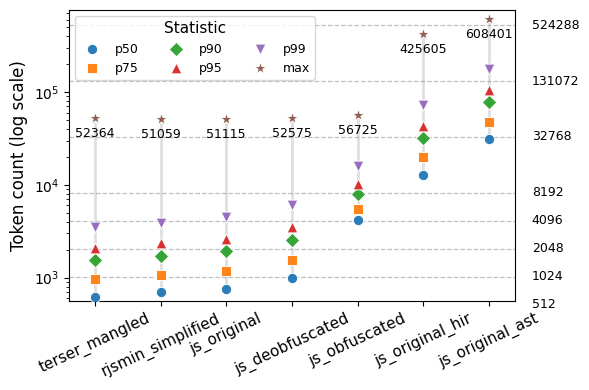}
    \caption{Tokenized Dataset Statistics, where PX marks that X percentage of samples fall below that token count.}
    \label{fig:tokens}
\end{figure}

Figure~\ref{fig:tokens} shows the distribution of tokenized sequence lengths in our dataset, which informs architecture choice: short to medium-length contexts fit well within encoder-only models.

In practice, trade-offs between context length, attention mechanism, and pretraining objectives determine how effectively stylistic signals are captured. Prior work~\cite{bisztray2025know} finds that representation quality can outweigh raw context window size. For example, a 512-token modified CodeT5 outperforms Longformer (which has a larger context-window).

\subsection{JavaScript Authorship Attribution}
Here, we present our key findings across different scenarios: (i) 5-class attribution, (ii) 10-class attribution, and (iii) 20-class attribution.

\subsubsection{Five-Class Attribution for Different Model Families}

We first replicate the setup of~\cite{bisztray2025know}, where the base BERT model achieved an accuracy of~$0.92$ on a five-way attribution task assigning C program code to one of five generator LLMs from distinct model families. 

Our selected models are \texttt{gpt-5-mini}, \texttt{gemini-2.5-flash-lite}, \texttt{qwen-2.5-coder-32b}, \texttt{llama-3.3-70b}, and \texttt{deepseek-v3.1}. This dataset consists of $12{,}500$ JavaScript programs ($5\times2{,}500$), split into 80\% training and 20\% validation. In this setting, BERT$_{B}$ achieved $97.9\%$ accuracy, F1, and precision/recall, confirming that LLMs from different families exhibit distinct and learnable stylistic signatures in JavaScript code generation, just like in C. Since we already achieved very high accuracy in this scenario, we shifted our focus to a more challenging problem, where all five models belong to the same family.

\subsubsection{Five-Class Attribution for Same Model Families}
  
Next, we test five-class attribution with GPT variants only. Here, BERT$_{B}$ still attains 90.2\% accuracy, indicating that even closely related LLMs leave detectable stylistic fingerprints. For a complete comparison of model performance on this task, see the summary Table~\ref{tab:three_variants}. Because we observe an approximate 8\% drop in accuracy compared to the different model family classification task, we evaluated a range of models to determine the best achievable performance.

\begin{figure}[b]
    \centering
    \includegraphics[width=1\linewidth]{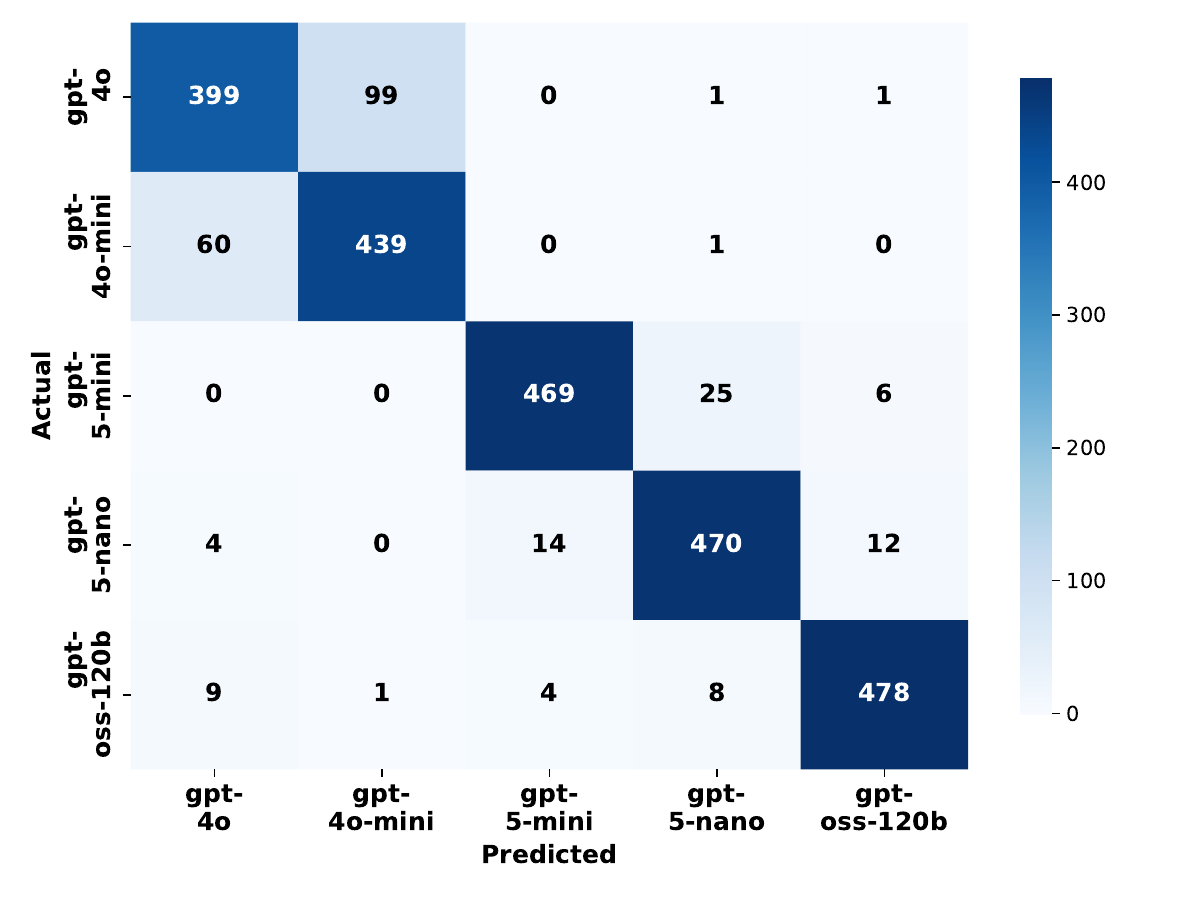}
    \caption{Confusion matrix for five-class attribution within the GPT family using BERT$_B$ on js\_original}
    \Description{Confusion matrix plot showing results for the same model families using BERT base on the original JavaScript source codes}
    \label{fig:same}
\end{figure}

Figure~\ref{fig:same} suggests that most errors arise from confusion between closely related variants of the same release (like GPT-4o vs. GPT-4o-mini, or among GPT-5 variants), while architecturally distinct versions can be distinguished more easily.

\definecolor{cyanblue}{RGB}{224,238,255}
\begin{table*}[ht]
\centering
\footnotesize
\caption{5/10/20 class attribution by JavaScript Variant (different model families)}
\renewcommand{\arraystretch}{1.0}
\rowcolors{2}{white}{cyanblue!70} 
\begin{tabular}{l| c| cccc| cccc| cccc| l}
\toprule
\multirow{2}{*}{\textbf{Model Name}} & \multirow{2}{*}{\textbf{Type}}
& \multicolumn{4}{c}{\textbf{5-Class (GPT family only)}}
& \multicolumn{4}{c}{\textbf{10-Class}}
& \multicolumn{4}{c}{\textbf{20-Class}}
& \multirow{2}{*}{\textbf{Parameters}} \\
\cmidrule(lr){3-6} \cmidrule(lr){7-10} \cmidrule(lr){11-14}
& & Acc(\%) & Prec(\%) & F1 & Time & Acc(\%) & Prec(\%) & F1 & Time & Acc(\%) & Prec(\%) & F1 & Time & \\ 
\midrule

\rowcolor{gray!35}
\multicolumn{15}{c}{\textbf{ORIGINAL JAVASCRIPT RESULTS}} \\

CodeT5-JSA        & LLM & 95.76 & 95.89 & 95.76 & 28:15 & 93.60 & 93.76 & 93.58 & 55:58 & 85.95 & 86.52 & 85:68 & 111:29 & Layers:24, Token:512 \\
Longformer$_{B}$  & LLM & 94.28 & 94.38 & 94.28 & 127:23 & 93.62 & 92.79 & 92.65 & 262:25 & 86.79 & 86.95 & 86.69 & 507:52 & Layers:12, Token:2048 \\
CodeBERT         & LLM & 94.28 & 94.40 & 94.27 & 07:00 & 91.88 & 92.11 & 91.86 & 13:15 & 85.87 & 86.22 & 85.74 & 25:08 & Layers:12, Token:512 \\
BERT$_{B}$       & LLM & 90.24 & 90.75 & 90.21 & 06:54 & 89.56 & 89.76 & 89.55 & 13:07 & 81.99 & 82.35 & 81.93 & 24:55 & Layers:12, Token:512 \\
\hline
XGBoost          & ML  & 88.72 & 88.76 & 88.74 & 00:49 & 86.60 & 86.67 & 86.46 & 02:23 & 75.78 & 75.73 & 75.62 & 08:31 & Estimators: 400 \\
Random Forest    & ML  & 86.24 & 86.27 & 86.25 & 00:35 & 81.20 & 81.85 & 80.75 & 01:15 & 68.25 & 68.28 & 67.19 & 02:56 & Trees: 400 \\
Linear SVM       & ML  & 82.64 & 82.60 & 82.60 & 00:01 & 76.44 & 75.83 & 75.83 & 00:02 & 58.97 & 58.17 & 57.73 & 00:12 & Max Iteration: 2000 \\
Logistic Regression & ML & 80.40 & 80.44 & 80.40 & 00:10 & 73.82 & 73.19 & 73.33 & 00:14 & 56.43 & 55.61 & 55.71 & 00:15 & Max Iteration: 2000 \\
KNN              & ML  & 71.32 & 71.41 & 71.21 & 00:01 & 60.74 & 61.49 & 59.99 & 00:01 & 48.03 & 49.70 & 46.34 & 00:01 & Neighbors: 5 \\
\specialrule{0.2em}{0pt}{0pt}

\rowcolor{gray!35}
\multicolumn{15}{c}{\textbf{MANGLED JAVASCRIPT RESULTS}} \\

CodeT5-JSA      & LLM & 90.02 & 90.2 & 90.03 & 28:16 & 89.68 & 89.85 & 89.64 & 55:58 & 83.61 & 83.74 & 83.41 & 112:02 & Layers:24, Token:512\\
Longformer$_{B}$ & LLM & 89.56 & 89.67 & 89.56 & 127:46 & 89.58 & 89.87 & 89.69 & 253:41 & 80.78 & 81.02 & 80.56 & 507:23 & Layers:12, Token:2048 \\
CodeBERT         & LLM & 89.28 & 89.54 & 89.26 & 06:55 & 89.36 & 89.89 & 89.40 & 13:00 & 81.28 & 81.49 & 81.13 & 25:26 & Layers:12, Token:512 \\
BERT$_{B}$       & LLM & 86.44 & 86.81 & 86.40 & 06:34 & 86.24 & 86.46 & 86.30 & 12:58 & 76.61 & 76.70 & 76.47 & 25:07 & Layers:12, Token:512 \\
\hline
XGBoost          & ML  & 86.48 & 86.51 & 86.49 & 00:51 & 84.80 & 84.87 & 84.68 & 150.54 & 73.29 & 73.08 & 73.04 & 595.40 & Estimators: 400 \\
Random Forest    & ML  & 83.04 & 83.05 & 82.99 & 00:31 & 79.72 & 80.55 & 79.31 & 68.30 & 66.93 & 66.88 & 65.98 & 168.78 & Trees: 400 \\
Linear SVM       & ML  & 79.08 & 79.03 & 78.97 & 00:01 & 74.78 & 74.59 & 74.31 & 2.67 & 55.77 & 54.79 & 54.27 & 15.28 & Max Iteration: 2000 \\
Logistic Regression & ML & 76.88 & 76.80 & 76.79 & 00:11 & 71.60 & 71.27 & 71.17 & 16.60 & 53.68 & 52.82 & 52.91 & 21.45 &  Max Iteration: 2000 \\
KNN              & ML  & 67.84 & 67.92 & 67.55 & 00:00 & 57.74 & 58.77 & 56.83 & 0.01 & 45.39 & 45.91 & 43.20 & 0.01 & Neighbors: 5 \\
\specialrule{0.2em}{0pt}{0pt}

\rowcolor{gray!35}
\multicolumn{15}{c}{\textbf{DEOBFUSCATED JAVASCRIPT RESULTS}} \\

CodeT5-JSA       & LLM & 89.92 & 90.40 & 89.86 & 28:12 & 89.78 & 90.00 & 89.78 & 56:08 & 79.94 & 79.79 & 79.60 & 112:44 & Layers:24, Token:512 \\
Longformer$_{B}$ & LLM & 88.32 & 88.42 & 88.32 & 127:14 & 88. 76& 89.14 & 88.84 & 254:13 & 78.15 & 78.02 & 77.87 & 508:35 & Layers:12, Token:2048 \\
CodeBERT         & LLM & 86.76 & 86.78 & 86.74 & 06:41 & 84.50 & 84.60 & 84.33 & 13:17 & 75.46 & 75.33 & 75.03 & 25:16 & Layers:12, Token:512 \\
BERT$_{B}$       & LLM & 84.08 & 84.26 & 84.08 & 06:55 & 82.58 & 82.84 & 82.63 & 12:34 & 72.61 & 72.64 & 72.40 & 24:44 & Layers:12, Token:512 \\
\hline
XGBoost          & ML  & 86.48 & 86.53 & 86.49 & 00:51 & 83.16 & 83.28 & 83.00 & 156.11 & 72.98 & 72.88 & 72.80 & 504.17 & Estimators: 400 \\
Random Forest    & ML  & 83.44 & 83.47 & 83.41 & 00:30 & 79.22 & 80.06 & 78.78 & 65.69 & 66.35 & 66.51 & 65.42 & 166.82 & Trees: 400 \\
Linear SVM       & ML  & 78.64 & 78.58 & 78.52 & 00:01 & 73.56 & 73.34 & 73.03 & 2.63 & 54.76 & 53.39 & 53.12 & 13.91 &  Max Iteration: 200 \\
Logistic Regression & ML & 77.76 & 77.72 & 77.67 & 00:08 & 71.40 & 71.10 & 70.96 & 14.59 & 53.43 & 52.44 & 52.66 & 19.42 &  Max Iteration: 2000 \\
KNN              & ML  & 70.28 & 70.30 & 70.06 & 00:00 & 60.40 & 61.47 & 59.64 & 0.01 & 47.63 & 48.05 & 45.63 & 0.01 & Neighbors: 5 \\
\specialrule{0.2em}{0pt}{0pt}

\bottomrule
\end{tabular}
\label{tab:three_variants}
\end{table*} 
Variants of the same model likely share similar pre-training corpora and compute budgets. We \emph{hypothesize} that their \textit{Supervised Fine-Tuning} (SFT) and \textit{Reinforcement Learning from Human Feedback} (RLHF) pipelines are also largely aligned. For example, GPT-4o and GPT-4o-mini may have been trained primarily on mid-2023 data, while GPT-5 variants likely incorporate early-2025 GitHub snapshots and more diverse coding examples, along with refined SFT and RLHF stages.

As AI-generated code becomes increasingly widespread, it remains uncertain whether model styles will converge into a uniform pattern or remain distinguishable through continued human-engineering interventions. We predict that as systems approach AGI, models will co-adapt and self-optimize, gradually blurring stylistic boundaries. Nonetheless, architectural and tokenization differences are expected to preserve some degree of distinguishability, if not through code style, then through higher-level moral, ethical, or policy biases imparted by their creators.

\subsubsection{Ten and Twenty-Class Attribution}
As the attribution task expands from five to ten and twenty classes, overall accuracy decreases across all models. Transformer-based encoders (\texttt{CodeT5-JSA}, \texttt{Longformer$_B$}, \texttt{CodeBERT}) consistently outperform classical ML baselines across all settings.

Among all models, \texttt{CodeT5-JSA} shows the best overall performance, maintaining 94\% accuracy in the ten-class setup and above 88\% in the twenty-class case for the original JavaScript. \texttt{Longformer$_B$} achieves comparable results but at substantially higher computational cost. Traditional ML classifiers such as \texttt{XGBoost} and \texttt{Random Forest} remain competitive yet degrade more rapidly as class count increases, and linear methods (\texttt{SVM}, \texttt{LogReg}) fall below 60\% at twenty classes. \texttt{XGBoost} consistently outperformed other ML models while remaining efficient, illustrating a favorable balance between accuracy and runtime. Surprisingly, on mangled and deobfuscated JavaScript code \texttt{XGBoost} managed to outperform the \texttt{Bert$_B$} model on several occasions.  

\subsubsection{Different JavaScript Variants}
To assess robustness, we evaluate attribution on terser-mangled and deobfuscated code (Table~\ref{tab:three_variants}). Since performance on mangled and minified JavaScript was nearly identical, we report results for the mangled variant only. Although these transformations eliminate key stylistic cues—such as variable and function names—accuracy remains high. This indicates that surface-level features like naming patterns or token frequencies are not the primary drivers of attribution. Instead, the classifiers exploit deeper structural and semantic regularities that persist through code transformations. When code is deobfuscated, accuracy recovers substantially, confirming that restoring syntactic structure is sufficient to reestablish most discriminative patterns.

These results extend the observations of~\cite{bisztray2025know} from C to JavaScript, demonstrating that attribution remains feasible both when the number of models scale, and with different transformations. Even as lexical information is deliberately suppressed, model-specific signatures persist through structural and dataflow patterns. Overall, these findings confirm that attribution grounded in semantic and syntactic organization remains tractable and robust even under aggressive obfuscation and fine-grained class expansion.

\subsection{Cross-check Validation}
All JavaScript samples in the \texttt{LLM-NodeJS} dataset were generated through OpenRouter.ai. To ensure that this service does not watermark its outputs, we created a new dataset of 500 JavaScript samples corresponding to completely unseen tasks. These samples, which include Node.js–specific programs, were generated using the original OpenAI API—entirely independent from the OpenRouter.ai service—and do not overlap with the 50,000 samples used for training or evaluation.

We re-evaluated the five-class classifier using \texttt{CodeBERT} on the original JavaScript variant. It maintained high attribution performance, with only a 1–2\% drop compared to the results observed in our experiments. This confirms that attribution generalizes well to previously unseen code distributions. The validation dataset is publicly released as \texttt{CROSS\_CHECK\_DATASET.json} on GitHub.

\section{Limitations and Future Work}
\label{sec:limitations}

This study leaves several open questions that point to promising directions for future research:
\begin{itemize}

\item  \textbf{Model scale and architectures:} Our experiments were limited to moderate-sized Transformer encoders such as BERT and CodeT5. Larger or extended-context architectures capable of processing JSIR and AST variants were not explored due to computational constraints. Future work should examine how model size, context length, and decoder-only architectures affect attribution accuracy. We did not conduct experiments on AST and JSIR representations, but made these variants available for the research community.

\item  \textbf{Language and data diversity:} The dataset is restricted to JavaScript, whereas prior work examined C~\cite{bisztray2025know}. Extending attribution to other languages (C++, Rust, Python, Java) or to cross-language setups---training on one language and testing on another---would help determine the generality of model fingerprints. Broader datasets covering varied coding domains and prompt styles would also strengthen external validity, as we only focused on web-deveopment.
    
\item \textbf{Attribution generalization:} In the multi-class setting, the classifier assumes that all samples originate from the $20$ studied LLMs. Future work should explore open-set recognition and continual learning to handle unseen models, as well as temporal drift studies to assess the stability of fingerprints as LLMs evolve.

\item \textbf{Interpretability and robustness.} Attribution robustness should be tested not only under simple code transformations, but under adversarial modifications like transpilation, control-flow rewriting.

\item \textbf{Ethical and practical considerations:} Attribution systems may be vulnerable to deliberate ``style spoofing'' or adversarial evasion. Future work should investigate these risks and develop safeguards for responsible use in code forensics and AI provenance detection.

\end{itemize}

\section{Conclusion}
\label{sec:conclusion}

We introduced the \texttt{LLM-NodeJS} dataset, a corpus of $250{,}000$ validated Node.js programs comprising $50{,}000$ original JavaScript programs, each with four variants (minified, terser-mangled, obfuscated, and deobfuscated), along with additional AST and JSIR representations for every sample. Using this dataset, we compared classical ML algorithms with fine-tuned Transformer encoders (e.g., BERT, CodeBERT, and Longformer) as well as a custom modified architecture based on the 770M-parameter CodeT5 model.

\begin{rqbox1}
\begin{itemize}
    \item \textbf{RQ1: Can we reliably attribute JavaScript programs to their generating LLM?} \\
    \textbf{Answer: Yes.} On original JavaScript, \textsc{CodeT5-JSA} attains \textbf{95.8\%} accuracy in the five-way GPT-only setting, \textbf{94.6\%} on the ten-class task, and \textbf{88.5\%} on twenty classes (Table~\ref{tab:three_variants}). Performance on terser-mangled, minified, and deobfuscated code remains close (typically single-digit drops), indicating that attribution does not depend primarily on surface naming or formatting.
\end{itemize}
\end{rqbox1}

\begin{rqbox1}
\begin{itemize}
    \item \textbf{RQ2: What signals do the models rely on?} \\
    \textbf{Answer: Deeper structural and semantic cues are just as important as style.} Our similarity analysis shows that \emph{dataflow} provides the clearest same-vs.-cross model separation, along with \emph{n-gram}, however, the absence of the latter still allows accurate attribution. These explain the robustness of attribution on mangled/minified code and the recovery after deobfuscation: classifiers exploit persistent logic, dependency structure, and control/dataflow rather than superficial lexical features.
\end{itemize}
\end{rqbox1}
Overall, our results demonstrate that LLM-generated JavaScript carries persistent, model-specific fingerprints that remain detectable across families, class scales (ten and twenty classes), and common code transformations. While accuracy naturally declines as stylistic boundaries narrow, semantically grounded signals continue to support reliable attribution.

\bibliographystyle{ACM-Reference-Format}

\bibliography{references}

@article{authorship,
author = {Juola, Patrick},
title = {Authorship attribution},
year = {2006},
issue_date = {December 2006},
publisher = {Now Publishers Inc.},
address = {Hanover, MA, USA},
volume = {1},
number = {3},
issn = {1554-0669},
url = {https://doi.org/10.1561/1500000005},
doi = {10.1561/1500000005},
journal = {Found. Trends Inf. Retr.},
month = dec,
pages = {233–334},
numpages = {102}
}

@inproceedings{10.1145/3605770.3625215,
author = {Bukhari, Sufiyan and Tan, Benjamin and De Carli, Lorenzo},
title = {Distinguishing AI- and Human-Generated Code: A Case Study},
year = {2023},
isbn = {9798400702631},
publisher = {Association for Computing Machinery},
address = {New York, NY, USA},
url = {https://doi.org/10.1145/3605770.3625215},
doi = {10.1145/3605770.3625215},
booktitle = {Proceedings of the 2023 Workshop on Software Supply Chain Offensive Research and Ecosystem Defenses},
pages = {17–25},
numpages = {9},
location = {Copenhagen, Denmark},
series = {SCORED '23}
}

@inproceedings{idialu2024whodunit,
  title={Whodunit: Classifying Code as Human Authored or GPT-4 generated-A case study on CodeChef problems},
  author={Idialu, Oseremen Joy and Mathews, Noble Saji and Maipradit, Rungroj and Atlee, Joanne M and Nagappan, Mei},
  booktitle={Proceedings of the 21st International Conference on Mining Software Repositories},
  pages={394--406},
  year={2024}
}

@inproceedings{simko2018,
  author       = {Lucy Simko and Luke Zettlemoyer and Tadayoshi Kohno},
  title        = {Recognizing and Imitating Programmer Style: Adversaries in Program Authorship Attribution},
  booktitle    = {Proceedings on Privacy Enhancing Technologies (PETS)},
  year         = {2018},
  volume       = {2018},
  number       = {1},
  pages        = {127--144},
  doi          = {10.1515/popets-2018-0007}
}

@inproceedings{mcknight2018,
  author       = {Christopher McKnight and Ian Goldberg},
  title        = {Style Counsel: Seeing the (Random) Forest for the Trees in Adversarial Code Stylometry},
  booktitle    = {ACM Workshop on Privacy in the Electronic Society (WPES)},
  year         = {2018},
  pages        = {138--142},
  publisher    = {ACM},
  address      = {Toronto, ON, Canada}
}

@article{Mitchell2023DataContamination,
  author    = {Melanie Mitchell},
  title     = {Data Contamination and Model Collapse in AI Systems},
  journal   = {AI Magazine},
  year      = {2023},
  volume    = {44},
  number    = {4},
  pages     = {358--364},
  doi       = {10.1002/aaai.12123},
}

@article{Shumailov2023CurseOfRecursion,
  author    = {Ilia Shumailov and Yarin Gal and Vitaly Shmatikov and Nicolas Papernot and Ross Anderson},
  title     = {The Curse of Recursion: Training on Generated Data Makes Models Forget},
  journal   = {Nature Machine Intelligence},
  year      = {2023},
  volume    = {5},
  pages     = {1411--1422},
  doi       = {10.1038/s42256-023-00782-5},
}

@misc{TheStack2022,
  author       = {{Kocetkov, Denis and Kumar, Divyanshu and Shoeb, Abu Awal Md and others}},
  title        = {The Stack: 3 TB of Permissively Licensed Source Code},
  howpublished = {Hugging Face Dataset},
  year         = {2022},
  url          = {https://huggingface.co/datasets/bigcode/the-stack},
  note         = {Accessed: 2025-09-09}
}

@misc{Amodei2025,
  author       = {Dario Amodei},
  title        = {Vibe Coding Is Coming for Engineering Jobs},
  howpublished = {\url{https://www.wired.com/story/vibe-coding-engineering-apocalypse}},
  year         = {2025},
  note         = {Accessed: 2025-09-09}
}

@misc{Nadella2025,
  author       = {Satya Nadella},
  title        = {As much as 30\% of Microsoft code now written by AI, CEO Satya Nadella says},
  howpublished = {\url{https://nypost.com/2025/04/30/business/microsoft-ceo-satya-nadella-says-30-of-code-now-written-by-ai}},
  year         = {2025},
  note         = {Accessed: 2025-09-09}
}

@misc{Altman2025,
  author       = {Sam Altman},
  title        = {Sam Altman says AI is doing 50\% of coding in companies},
  howpublished = {\url{https://www.indiatoday.in/technology/news/story/sam-altman-says-ai-is-doing-50-percent-coding-in-companies-warns-students-about-career-choices-2696937-2025-03-21}},
  year         = {2025},
  note         = {Accessed: 2025-09-09}
}

@inproceedings{OmanCook1989,
  author    = {Paul W. Oman and Curtis R. Cook},
  title     = {Programming Style Authorship Analysis},
  booktitle = {Proceedings of the 17th Annual ACM Computer Science Conference (CSC '89)},
  year      = {1989},
  pages     = {320--326},
  address   = {New York, NY, USA},
  publisher = {Association for Computing Machinery},
}

@InProceedings{10.1007/978-981-16-6372-7_70,
author="Guo, Dixiao
and Zhou, Anmin
and Liu, Liang
and Liao, Shan
and Zhang, Lei",
editor="Deng, Zhidong",
title="A Method of Source Code Authorship Attribution Based on Graph Neural Network",
booktitle="Proceedings of 2021 Chinese Intelligent Automation Conference",
year="2022",
publisher="Springer Singapore",
address="Singapore",
pages="645--657",
isbn="978-981-16-6372-7"
}

@inproceedings{bogomolov2021authorship,
  author    = {Egor Bogomolov and Vladimir Kovalenko and Yurii Rebryk and Alberto Bacchelli and Timofey Bryksin},
  title     = {Authorship Attribution of Source Code: A Language-Agnostic Approach and Applicability in Software Engineering},
  booktitle = {Proceedings of the 29th ACM Joint European Software Engineering Conference and Symposium on the Foundations of Software Engineering (ESEC/FSE ’21)},
  year      = {2021},
  pages     = {932--944},
  address   = {Athens, Greece},
  publisher = {Association for Computing Machinery},
  isbn      = {978-1-4503-8562-6},
  doi       = {10.1145/3468264.3468606},
}

@article{bisztray2025know,
  title={I Know Which LLM Wrote Your Code Last Summer: LLM generated Code Stylometry for Authorship Attribution},
  author={Bisztray, Tamas and Cherif, Bilel and Dubniczky, Richard A and Gruschka, Nils and Borsos, Bertalan and Ferrag, Mohamed Amine and Kovacs, Attila and Mavroeidis, Vasileios and Tihanyi, Norbert},
  journal={arXiv preprint arXiv:2506.17323},
  year={2025}
}

@Article{info15120819,
AUTHOR = {Bulla, Luana and Midolo, Alessandro and Mongiovì, Misael and Tramontana, Emiliano},
TITLE = {EX-CODE: A Robust and Explainable Model to Detect AI-Generated Code},
JOURNAL = {Information},
VOLUME = {15},
YEAR = {2024},
NUMBER = {12},
ARTICLE-NUMBER = {819},
URL = {https://www.mdpi.com/2078-2489/15/12/819},
ISSN = {2078-2489},
DOI = {10.3390/info15120819}
}

@inproceedings{vaswani_attention_2017,
author = {Vaswani, Ashish and Shazeer, Noam and Parmar, Niki and Uszkoreit, Jakob and Jones, Llion and Gomez, Aidan N. and Kaiser, \L{}ukasz and Polosukhin, Illia},
title = {Attention is all you need},
year = {2017},
isbn = {9781510860964},
publisher = {Curran Associates Inc.},
address = {Red Hook, NY, USA},
booktitle = {Proceedings of the 31st International Conference on Neural Information Processing Systems},
pages = {6000–6010},
numpages = {11},
location = {Long Beach, California, USA},
series = {NIPS'17}
}

@inproceedings{howard-ruder-2018-universal,
    title = "Universal Language Model Fine-tuning for Text Classification",
    author = "Howard, Jeremy  and
      Ruder, Sebastian",
    editor = "Gurevych, Iryna  and
      Miyao, Yusuke",
    booktitle = "Proceedings of the 56th Annual Meeting of the Association for Computational Linguistics (Volume 1: Long Papers)",
    month = jul,
    year = "2018",
    address = "Melbourne, Australia",
    publisher = "Association for Computational Linguistics",
    url = "https://aclanthology.org/P18-1031/",
    doi = "10.18653/v1/P18-1031",
    pages = "328--339"
}

@misc{wang_codet5_2021,
	title = {{CodeT5}: {Identifier}-aware {Unified} {Pre}-trained {Encoder}-{Decoder} {Models} for {Code} {Understanding} and {Generation}},
	shorttitle = {{CodeT5}},
	url = {http://arxiv.org/abs/2109.00859},
	doi = {10.48550/arXiv.2109.00859},
	urldate = {2025-06-11},
	publisher = {arXiv},
	author = {Wang, Yue and Wang, Weishi and Joty, Shafiq and Hoi, Steven C. H.},
	month = sep,
	year = {2021},
	note = {arXiv:2109.00859 [cs]},
}

@misc{guo_graphcodebert_2021,
	title = {{GraphCodeBERT}: {Pre}-training {Code} {Representations} with {Data} {Flow}},
	shorttitle = {{GraphCodeBERT}},
	url = {http://arxiv.org/abs/2009.08366},
	doi = {10.48550/arXiv.2009.08366},
	urldate = {2025-06-11},
	publisher = {arXiv},
	author = {Guo, Daya and Ren, Shuo and Lu, Shuai and Feng, Zhangyin and Tang, Duyu and Liu, Shujie and Zhou, Long and Duan, Nan and Svyatkovskiy, Alexey and Fu, Shengyu and Tufano, Michele and Deng, Shao Kun and Clement, Colin and Drain, Dawn and Sundaresan, Neel and Yin, Jian and Jiang, Daxin and Zhou, Ming},
	month = sep,
	year = {2021},
	note = {arXiv:2009.08366 [cs]},
}

@misc{feng_codebert_2020,
	title = {{CodeBERT}: {A} {Pre}-{Trained} {Model} for {Programming} and {Natural} {Languages}},
	shorttitle = {{CodeBERT}},
	url = {http://arxiv.org/abs/2002.08155},
	doi = {10.48550/arXiv.2002.08155},
	urldate = {2025-06-11},
	publisher = {arXiv},
	author = {Feng, Zhangyin and Guo, Daya and Tang, Duyu and Duan, Nan and Feng, Xiaocheng and Gong, Ming and Shou, Linjun and Qin, Bing and Liu, Ting and Jiang, Daxin and Zhou, Ming},
	month = sep,
	year = {2020},
	note = {arXiv:2002.08155 [cs]},
}

@INPROCEEDINGS{7784595,
  author={Ferrante, Alberto and Medvet, Eric and Mercaldo, Francesco and Milosevic, Jelena and Visaggio, Corrado Aaron},
  booktitle={2016 11th International Conference on Availability, Reliability and Security (ARES)}, 
  title={Spotting the Malicious Moment: Characterizing Malware Behavior Using Dynamic Features}, 
  year={2016},
  volume={},
  number={},
    publisher  = {IEEE},
  address    = {Salzburg, Austria},
  pages={372-381},
  doi={10.1109/ARES.2016.70}}

@INPROCEEDINGS {9825799,
author = { Song, Qige and Zhang, Yongzheng and Ouyang, Linshu and Chen, Yige },
booktitle = { 2022 IEEE International Conference on Software Analysis, Evolution and Reengineering (SANER) },
title = {{ BinMLM: Binary Authorship Verification with Flow-aware Mixture-of-Shared Language Model }},
year = {2022},
volume = {},
ISSN = {1534-5351},
pages = {1023-1033},
doi = {10.1109/SANER53432.2022.00120},
url = {https://doi.ieeecomputersociety.org/10.1109/SANER53432.2022.00120},
publisher = {IEEE Computer Society},
address = {Los Alamitos, CA, USA},
month =mar}

@article{10.1145/3653973,
author = {Gray, Jason and Sgandurra, Daniele and Cavallaro, Lorenzo and Blasco Alis, Jorge},
title = {Identifying Authorship in Malicious Binaries: Features, Challenges \& Datasets},
year = {2024},
issue_date = {August 2024},
publisher = {Association for Computing Machinery},
address = {New York, NY, USA},
volume = {56},
number = {8},
issn = {0360-0300},
url = {https://doi.org/10.1145/3653973},
doi = {10.1145/3653973},
journal = {ACM Comput. Surv.},
month = apr,
articleno = {212},
numpages = {36}
}

@article{nguyen_gptsniffer_2024,
	title = {{GPTSniffer}: {A} {CodeBERT}-based classifier to detect source code written by {ChatGPT}},
	volume = {214},
	issn = {0164-1212},
	shorttitle = {{GPTSniffer}},
	url = {https://www.sciencedirect.com/science/article/pii/S0164121224001043},
	doi = {10.1016/j.jss.2024.112059},
	urldate = {2025-06-01},
	journal = {Journal of Systems and Software},
	author = {Nguyen, Phuong T. and Di Rocco, Juri and Di Sipio, Claudio and Rubei, Riccardo and Di Ruscio, Davide and Di Penta, Massimiliano},
	month = aug,
	year = {2024},
	pages = {112059},
}

@inproceedings{pan2024assessing,
  title={Assessing ai detectors in identifying ai-generated code: Implications for education},
  author={Pan, Wei Hung and Chok, Ming Jie and Wong, Jonathan Leong Shan and Shin, Yung Xin and Poon, Yeong Shian and Yang, Zhou and Chong, Chun Yong and Lo, David and Lim, Mei Kuan},
  booktitle={Proceedings of the 46th international conference on software engineering: software engineering education and training},
  pages={1--11},
  year={2024}
}

@article{zafar_language_2020,
	title = {Language and {Obfuscation} {Oblivious} {Source} {Code} {Authorship} {Attribution}},
	volume = {8},
	issn = {2169-3536},
	url = {https://ieeexplore.ieee.org/abstract/document/9245552},
	doi = {10.1109/ACCESS.2020.3034932},
	urldate = {2025-05-31},
	journal = {IEEE Access},
	author = {Zafar, Sarim and Sarwar, Muhammad Usman and Salem, Saeed and Malik, Muhammad Zubair},
	year = {2020},
	pages = {197581--197596},
}

@inproceedings{rosenblum_who_2011,
	address = {Berlin, Heidelberg},
	title = {Who {Wrote} {This} {Code}? {Identifying} the {Authors} of {Program} {Binaries}},
	isbn = {978-3-642-23822-2},
	shorttitle = {Who {Wrote} {This} {Code}?},
	doi = {10.1007/978-3-642-23822-2_10},
	language = {en},
	booktitle = {Computer {Security} – {ESORICS} 2011},
	publisher = {Springer},
	author = {Rosenblum, Nathan and Zhu, Xiaojin and Miller, Barton P.},
	editor = {Atluri, Vijay and Diaz, Claudia},
	year = {2011},
	pages = {172--189},
}

@inproceedings{guo_method_2022,
	address = {Singapore},
	title = {A {Method} of {Source} {Code} {Authorship} {Attribution} {Based} on {Graph} {Neural} {Network}},
	isbn = {978-981-16-6372-7},
	doi = {10.1007/978-981-16-6372-7_70},
	language = {en},
	booktitle = {Proceedings of 2021 {Chinese} {Intelligent} {Automation} {Conference}},
	publisher = {Springer},
	author = {Guo, Dixiao and Zhou, Anmin and Liu, Liang and Liao, Shan and Zhang, Lei},
	editor = {Deng, Zhidong},
	year = {2022},
	pages = {645--657},
}

@article{abuhamad_large-scale_2021,
	title = {Large-scale and {Robust} {Code} {Authorship} {Identification} with {Deep} {Feature} {Learning}},
	volume = {24},
	issn = {2471-2566},
	url = {https://dl.acm.org/doi/10.1145/3461666},
	doi = {10.1145/3461666},
	number = {4},
	urldate = {2025-05-30},
	journal = {ACM Trans. Priv. Secur.},
	author = {Abuhamad, Mohammed and Abuhmed, Tamer and Mohaisen, David and Nyang, Daehun},
	month = jul,
	year = {2021},
	pages = {23:1--23:35},
}

@inproceedings{alsulami_source_2017,
	address = {Cham},
	title = {Source {Code} {Authorship} {Attribution} {Using} {Long} {Short}-{Term} {Memory} {Based} {Networks}},
	isbn = {978-3-319-66402-6},
	doi = {10.1007/978-3-319-66402-6_6},
	language = {en},
	booktitle = {Computer {Security} – {ESORICS} 2017},
	publisher = {Springer International Publishing},
	author = {Alsulami, Bander and Dauber, Edwin and Harang, Richard and Mancoridis, Spiros and Greenstadt, Rachel},
	editor = {Foley, Simon N. and Gollmann, Dieter and Snekkenes, Einar},
	year = {2017},
	pages = {65--82},
}

@inproceedings{wang_integration_2018,
	address = {New York, NY, USA},
	series = {{AISec} '18},
	title = {Integration of {Static} and {Dynamic} {Code} {Stylometry} {Analysis} for {Programmer} {De}-anonymization},
	isbn = {978-1-4503-6004-3},
	url = {https://dl.acm.org/doi/10.1145/3270101.3270110},
	doi = {10.1145/3270101.3270110},
	urldate = {2025-05-30},
	booktitle = {Proceedings of the 11th {ACM} {Workshop} on {Artificial} {Intelligence} and {Security}},
	publisher = {Association for Computing Machinery},
	author = {Wang, Ningfei and Ji, Shouling and Wang, Ting},
	month = jan,
	year = {2018},
	pages = {74--84},
}

@inproceedings{caliskan_when_2018,
	address = {San Diego, CA},
	title = {When {Coding} {Style} {Survives} {Compilation}: {De}-anonymizing {Programmers} from {Executable} {Binaries}},
	isbn = {978-1-891562-49-5},
	shorttitle = {When {Coding} {Style} {Survives} {Compilation}},
	url = {https://www.ndss-symposium.org/wp-content/uploads/2018/02/ndss2018_06B-2_Caliskan_paper.pdf},
	doi = {10.14722/ndss.2018.23304},
	language = {en},
    numpages = {15},
	urldate = {2025-05-30},
	booktitle = {Proceedings 2018 {Network} and {Distributed} {System} {Security} {Symposium}},
	publisher = {Internet Society},
	author = {Caliskan, Aylin and Yamaguchi, Fabian and Dauber, Edwin and Harang, Richard and Rieck, Konrad and Greenstadt, Rachel and Narayanan, Arvind},
	year = {2018},
}

@inproceedings{quiring_misleading_2019,
author = {Quiring, Erwin and Maier, Alwin and Rieck, Konrad},
title = {Misleading authorship attribution of source code using adversarial learning},
year = {2019},
isbn = {9781939133069},
publisher = {USENIX Association},
address = {USA},
booktitle = {Proceedings of the 28th USENIX Conference on Security Symposium},
pages = {479–496},
numpages = {18},
location = {Santa Clara, CA, USA},
series = {SEC'19}
}

@inproceedings{caliskan-islam_-anonymizing_2015,
author = {Caliskan-Islam, Aylin and Harang, Richard and Liu, Andrew and Narayanan, Arvind and Voss, Clare and Yamaguchi, Fabian and Greenstadt, Rachel},
title = {De-anonymizing programmers via code stylometry},
year = {2015},
isbn = {9781931971232},
publisher = {USENIX Association},
address = {USA},
booktitle = {Proceedings of the 24th USENIX Conference on Security Symposium},
pages = {255–270},
numpages = {16},
location = {Washington, D.C.},
series = {SEC'15}
}

@misc{huang_authorship_2025,
	title = {Authorship {Attribution} in the {Era} of {LLMs}: {Problems}, {Methodologies}, and {Challenges}},
	shorttitle = {Authorship {Attribution} in the {Era} of {LLMs}},
	url = {http://arxiv.org/abs/2408.08946},
	doi = {10.48550/arXiv.2408.08946},
	urldate = {2025-05-29},
	publisher = {arXiv},
	author = {Huang, Baixiang and Chen, Canyu and Shu, Kai},
	month = jan,
	year = {2025},
	note = {arXiv:2408.08946 [cs]
version: 2},
}

@misc{rahman_automatic_2024,
	title = {Automatic {Detection} of {LLM}-generated {Code}: {A} {Case} {Study} of {Claude} 3 {Haiku}},
	shorttitle = {Automatic {Detection} of {LLM}-generated {Code}},
	url = {http://arxiv.org/abs/2409.01382},
	doi = {10.48550/arXiv.2409.01382},
	urldate = {2025-05-29},
	publisher = {arXiv},
	author = {Rahman, Musfiqur and Khatoonabadi, SayedHassan and Abdellatif, Ahmad and Shihab, Emad},
	month = sep,
	year = {2024},
	note = {arXiv:2409.01382 [cs]
version: 1},
}

@INPROCEEDINGS {10992332,
author = { Gurioli, Andrea and Gabbrielli, Maurizio and Zacchiroli, Stefano },
booktitle = { 2025 IEEE International Conference on Software Analysis, Evolution and Reengineering (SANER) },
title = {{ Is This You, LLM? Recognizing AI-written Programs with Multilingual Code Stylometry }},
year = {2025},
volume = {},
ISSN = {},
pages = {394-405},
doi = {10.1109/SANER64311.2025.00044},
url = {https://doi.ieeecomputersociety.org/10.1109/SANER64311.2025.00044},
publisher = {IEEE Computer Society},
address = {Los Alamitos, CA, USA},
month =mar}

@misc{choi_i_2025,
	title = {I {Can} {Find} {You} in {Seconds}! {Leveraging} {Large} {Language} {Models} for {Code} {Authorship} {Attribution}},
	url = {http://arxiv.org/abs/2501.08165},
	doi = {10.48550/arXiv.2501.08165},
	urldate = {2025-05-29},
	publisher = {arXiv},
	author = {Choi, Soohyeon and Tan, Yong Kiam and Meng, Mark Huasong and Ragab, Mohamed and Mondal, Soumik and Mohaisen, David and Aung, Khin Mi Mi},
	month = jan,
	year = {2025},
	note = {arXiv:2501.08165 [cs]
version: 1},
}

@article{tihanyi_how_2024,
	title = {How secure is {AI}-generated code: a large-scale comparison of large language models},
	volume = {30},
	issn = {1573-7616},
	shorttitle = {How secure is {AI}-generated code},
	url = {https://doi.org/10.1007/s10664-024-10590-1},
	doi = {10.1007/s10664-024-10590-1},
	language = {en},
	number = {2},
	urldate = {2025-05-29},
	journal = {Empirical Software Engineering},
	author = {Tihanyi, Norbert and Bisztray, Tamas and Ferrag, Mohamed Amine and Jain, Ridhi and Cordeiro, Lucas C.},
	month = dec,
	year = {2024},
	pages = {47},
}

@article{he_authorship_2024,
	title = {Authorship {Attribution} {Methods}, {Challenges}, and {Future} {Research} {Directions}: {A} {Comprehensive} {Survey}},
	volume = {15},
	copyright = {http://creativecommons.org/licenses/by/3.0/},
	issn = {2078-2489},
	shorttitle = {Authorship {Attribution} {Methods}, {Challenges}, and {Future} {Research} {Directions}},
	url = {https://www.mdpi.com/2078-2489/15/3/131},
	doi = {10.3390/info15030131},
	language = {en},
	number = {3},
	urldate = {2025-05-28},
	journal = {Information},
	author = {He, Xie and Lashkari, Arash Habibi and Vombatkere, Nikhill and Sharma, Dilli Prasad},
	month = mar,
	year = {2024},
	note = {Number: 3
Publisher: Multidisciplinary Digital Publishing Institute},
	pages = {131},
}

@String{Computing = "Computing" }

@String{Computer = "{IEEE} Computer" }

@String{Springer = "Springer-Verlag" }

@inproceedings{abuhamad_large-scale_2018,
    address = {New York, NY, USA},
    series = {{CCS} '18},
    title = {Large-{Scale} and {Language}-{Oblivious} {Code} {Authorship} {Identification}},
    isbn = {978-1-4503-5693-0},
    url = {https://dl.acm.org/doi/10.1145/3243734.3243738},
    doi = {10.1145/3243734.3243738},
    urldate = {2025-05-31},
    booktitle = {Proceedings of the 2018 {ACM} {SIGSAC} {Conference} on {Computer} and {Communications} {Security}},
    publisher = {Association for Computing Machinery},
    author = {Abuhamad, Mohammed and AbuHmed, Tamer and Mohaisen, Aziz and Nyang, DaeHun},
    month = oct,
    year = {2018},
    pages = {101--114},
}

\end{document}